# NeRF View Synthesis: Subjective Quality Assessment and Objective Metrics Evaluation

**Pedro Martin, António Rodrigues, João Ascenso, and Maria Paula Queluz**
Instituto de Telecomunicações/Instituto Superior Técnico, University of Lisbon, 1049-001 Lisbon, Portugal

Corresponding author: Pedro Martin (e-mail: pedro.martin@lx.it.pt).

This work was supported by the Fundação para a Ciencia e a Tecnologia (FCT) through national funds, and when applicable, co-funded by EU funds under the projects UIDB/50008/2020 and PTDC/EEI-COM/7775/2020.

**ABSTRACT** Neural radiance fields (NeRF) are a groundbreaking computer vision technology that enables the generation of high-quality, immersive visual content from multiple viewpoints. This capability has significant advantages for applications such as virtual/augmented reality, 3D modelling, and content creation for the film and entertainment industry. However, the evaluation of NeRF methods poses several challenges, including a lack of comprehensive datasets, reliable assessment methodologies, and objective quality metrics. This paper addresses the problem of NeRF view synthesis (NVS) quality assessment thoroughly, by conducting a rigorous subjective quality assessment test that considers several scene classes and recently proposed NVS methods. Additionally, the performance of a wide range of state-of-the-art conventional and learning-based full-reference 2D image and video quality assessment metrics is evaluated against the subjective scores of the subjective study. This study found that errors in camera pose estimation can result in spatial misalignments between synthesized and reference images, which need to be corrected before applying an objective quality metric. The experimental results are analyzed in depth, providing a comparative evaluation of several NVS methods and objective quality metrics, across different classes of visual scenes, including real and synthetic content for front-face and 360º camera trajectories.

## I. INTRODUCTION

Nowadays, 3D visual representations are enabling more immersive and interactive multimedia experiences, not achievable with traditional 2D visual content. A wide range of virtual reality (VR) applications and services are now possible, including virtual shopping, virtual tours, remote education, and gaming, among others. To enhance user immersion through six degrees-of-freedom (6DoF) content navigation, visual representation methods such as multiview+depth (MVD), light fields (LF) and point clouds (PC) have been proposed in recent years. The chosen representation has a direct impact on the 6DoF system pipeline, notably on the rendering component; while a point cloud may be rendered directly using some elementary structure, MVD or LF requires the synthesis of novel views using, commonly, depth image-based rendering (DIBR) methods [1], [2].

Neural Radiance Fields (NeRF) have recently emerged as a promising solution for representing and rendering 3D scenes and have already achieved state-of-the-art results in view synthesis [3]. Due to their potential, NeRF methods are attracting significant attention from the research community from both academia and industry. Moreover, they quickly began to be used to solve adjacent problems, such as super-resolution, pose estimation, depth estimation, among others [4].

However, NeRF view synthesis (NVS) methods have their own structural limitations, which may lead to a variety of artifacts in the final visual result, with a negative impact on the perceived quality. Currently, the quality of NVS is assessed using 2D image and video quality assessment metrics (IQA and VQA), namely PSNR, SSIM [5], MS-SSIM [6], LPIPS [7], and FovVideoVDP [8]. However, there is no evidence about the quality assessment performance of these metrics in this context. In fact, NVS methods are known to generate new artifacts, notably the so-called floaters, besides flawed object geometry and flickering edges. To know the impact of these new artifacts in human perception, subjective assessment studies under controlled and well-known conditions are much needed. Thus, the main objective of this work is to study, in a subjective way, the impact of the different artifacts produced by several NVS methods for different classes of visual scenes (e.g. 360º), and to evaluate the performance of state-of-the-art IQA and VQA metrics considering the results obtained on the subjective assessment study. The subjective evaluation considers an extensive and relevant range of real and synthetic 3D scenes, with front-facing (FF) and 360º camera acquisitions, and several recently proposed NVS methods. The key contributions are:

- Creation of a new dataset of FF synthetic and real visual scenes with the respective camera poses, that can be

used to assess NVS methods. All visual scenes were acquired with a (virtual or real) camera and thus video sequences are available for full-reference evaluation of synthesized videos, allowing to compare NVS methods in a consistent way for FF and 360º visual scenes. This contribution is of great interest to the NVS community especially because video sequences are typically not available for FF scenes.

- Evaluation of the impact, on perceived quality, of NVS using a well-known and reliable subjective assessment methodology. In this study, several scene classes and recently proposed NVS methods are considered. This contribution is critical for the young NVS quality assessment community, as there is no such extensive subjective quality database. Moreover, this study allows to identify the strengths and weaknesses of several popular NVS solutions for different scene classes.
- Evaluation, in a NVS context, of objective quality assessment metrics developed for 2D images and video, using several scene classes (namely real and synthetic 360º and FF scenes). This contribution is fundamental for inferring whether the existing quality metrics enable accurate NVS quality assessments, or for getting insights into the approaches that should be pursued in the development of novel metrics, specifically designed for NVS. Additionally, this study found that errors in camera pose estimation can lead to spatial misalignments between reference and synthesized images, which should be corrected before applying an objective quality assessment metric.

All reference and synthesized video sequences, and the corresponding scores from the subjective assessment study, constitute the NVS-QA dataset that can be found at https://github.com/pedrogcmartin/NVS-QA. The remaining of this paper is organized as follows: Section II provides an overview of the related work. Section III presents the fundamental concepts, selected NVS methods, considered scene classes, and artifact characterization. Section IV details the subjective assessment study that was conducted. Section V presents the analysis that was carried out regarding the objective quality assessment metrics. Section VI presents the conclusions. Finally, Section VII discusses challenges and suggests potential ways forward toward the advancement of NVS quality assessment.

## II. RELATED WORK

In the past, several subjective and objective quality assessment studies have been performed for classical hand-crafted DIBR-based view synthesis [9]-[12] seeking the development of quality assessment databases and/or accurate objective quality metrics. More recently, some preliminary subjective studies for NVS methods were presented by us in [13], followed by other more recent works [14]-[17]. The development of objective quality metrics and evaluation frameworks, in the NVS context, have also been presented in [17], [18].

### A. DIBR QUALITY ASSESSMENT

Considering DIBR quality assessment, it was proposed in [9] the IRCCyN/IVC DIBR video quality database, containing 26 reference video sequences, 84 video sequences synthesized by seven DIBR algorithms, and the respective quality scores obtained from a subjective test campaign. The synthesized videos were also evaluated with conventional objective image and video quality metrics, and the resulting scores were compared with the human judgments. The results showed the inefficiency of the considered objective metrics to properly evaluate the quality of DIBR synthesized views. In [10], the impact of MVD texture and depth compression on the quality of DIBR synthesis was analyzed, based on a new video quality database (named as SIAT) consisting of 10 MVD reference sequences, 140 synthesized video sequences, and their respective subjective scores; the VSRS-1D-Fast algorithm was used for view synthesis. A novel full-reference objective metric was further proposed with focus on temporal flickering distortions caused by the depth compression. In [11], the authors proposed the IST view synthesis database containing 180 images synthesized with the VSRS-1D-Fast and the VSIM algorithms [19]-[21], the respective reference images, the texture and depth maps of the lateral source images used for view synthesis, and the subjective scores of the synthesized images. Conventional 2D full-reference metrics were evaluated by measuring the correlation with subjective scores. A support vector regression algorithm was trained with features extracted before, during, and after the synthesis procedure, and used to design a no-reference metric that outperforms the conventional metrics. However, as the proposed metric requires specific information from the DIBR process, it cannot be directly applied to NVS. In [12], the NBU-3D Synthesized Video Quality Database was proposed, with 128 synthesized videos (using the VSRS-1D-Fast) generated from several combinations of texture/depth compression, together with the corresponding subjective scores. Its main purpose was to train a new video quality metric that learned the relationship between features extracted from texture and depth maps and perceived video quality. NVS does not use depth maps, meaning that the proposed learning-based metric is not useful in this context.

### B. NVS QUALITY ASSESSMENT

The quality assessment of NVS has not received much attention from the scientific community. To the best of the authors knowledge, our previous work [13] is the first published and peer-reviewed contribution on the subject, where a subjective test campaign of videos synthesized with several NVS methods, and considering real and synthetic 360º scenes, was conducted; this was an initial study without FF video sequences and without the evaluation of any objective metric. In [14], the performance of a small set of quality metrics developed for 2D images and video, when applied to NVS, is evaluated using the reference and synthesized video sequences of the NeRF-QA database proposed by us in [13]. However, besides the small set of considered metrics, the conclusions of [14] were also limited by the absence of significance tests on the metrics results, as well as the lack of FF scenes. In [15], a subjective and objective quality evaluation of NVS methods was carried out; however, only FF real scenes, acquired with restricted camera motion and using a 2D gantry or a slider (that is hardly used in many real-world acquisitions) were considered. Recently, a new NVS database (named as Explicit-NeRF-QA) is proposed in [16], which includes 22 diverse 3D objects to train explicit NVS methods. Lossy compression was simulated by adjusting key parameters within the selected NVS methods (e.g. hash table size for Instant-NGP [22] and the voxel grid resolution for Plenoxels [23]) during NVS training. A subjective study was conducted with 21 participants and the performance of state-of-the-art objective quality metrics was analyzed. However, the database is strictly limited to bounded synthetic scenes generated by explicit NVS methods and no practical compression methods were used. A new NeRF evaluation framework is proposed in [18], which trains and evaluates NVS methods on an explicit radiance field representation. The mean absolute error between the learned NVS outputs and the corresponding ground truth lead to the creation of a novel objective metric. Despite having the benefits of considering the whole NVS pipeline, the proposed framework restricts NVS evaluation to synthetic scenes. In [17], the authors proposed an objective quality metric specifically created for NVS content, called as NeRF-NQA. It operates through three modules: the viewwise (analyzes spatial quality features across views); the pointwise (captures angular features based on views and camera poses); and the quality score estimation module (fuses both sets of features via an MLP to deliver a final quality score). However, the method was tested only on FF real scenes and on synthesized images obtained with more recent NVS methods (such as Plenoxels and Instant-NGP).

The quality assessment study presented in this paper includes FF scenes with higher camera motion than in [15], the use of FF synthetic content, and of 360º real and synthetic content. Furthermore, all the real scenes were captured using a standard camera equipped with simple stabilizer, within an uncontrolled environment, thereby being closer to a practical use case. The focus is to evaluate rendering artifacts, which are rather significant without explicitly considering compression artifacts as in [16]. A wide range of IQA and VQA metrics (including learning-based metrics) representative of the state-of-the-art, were also selected for NVS quality as-assessment. All observations and conclusions drawn from this study are supported by statistical tests of significance. Another key contribution is the use of spatial shift compensation in the experimental analysis. In fact, it was verified that the pose estimation process (used before the NVS training) may lead to significant displacements between reference and synthesized images, and that the performance of the objective quality metrics may be enhanced by correcting this shift before applying the metrics.

## III. NERF VIEW SYNTHESIS

This section describes the fundamentals behind NVS, presents the visual scene classes and presents the NVS methods that were selected for the quality assessment study, and characterizes the artifacts that are typically generated by the synthesis process.

### A. NERF VIEW SYNTHESIS FUNDAMENTALS

The NeRF framework was introduced in the seminal work of [3]. Its breakthrough lies in representing the visual scene as a continuous 5D function that encodes both spatial location and viewing direction. In [3], this 5D function is a multi-layer perceptron (MLP) which is able to learn the correspondence between the spatial location of 3D scene points located along a given viewing direction, and its respective color and opacity level. To perform view synthesis, the MLP is directly queried for a set of 3D points sampled along the camera ray corresponding to the pixel to be rendered, followed by classic volume rendering technique applied to the query results. The MLP is trained by successively comparing the synthesized image with a reference training image, for a set of different viewpoints. As such, the only information needed by a NVS method to learn a given visual scene is a set of acquired 2D images of that scene, and the respective camera poses, without explicitly reconstructing the 3D scene geometry.

Since the seminal NeRF struggles to obtain high quality synthesis, especially for real scenes, several authors have proposed improvements to it (e.g., [24], [25]), while keeping the use of MLPs in the training process. Additionally, due to the high training and synthesis times of MLPs, a branch of MLP-free NeRF-based methods was also proposed [4]; these methods adopt an explicit scene representation through 3D voxel grids, enabling direct estimation of the scene geometry and luminosity characteristics.

### B. VISUAL SCENE CLASSES

Visual scenes are organized into classes which are described next along with their main characteristics:

- **Synthetic or real content:** Synthetic content is computer generated imagery and can represent highly realistic objects, scenes with complex geometries, and non-Lambertian materials. Real content is acquired from real-world scenes, using a camera system with an optical sensor.
- **360º or front-facing content:** In 360º synthetic scenes, a set of virtual cameras is placed at the same distance from the scene (or object) center, forming a semi-sphere or a full-sphere constellation of inward-facing cameras (Fig. 1-a)). In 360º real scenes, the camera moves around a given region of interest, typically approaching a circular inward-pointing movement (Fig. 1-b)). Both real and synthetic FF scenes are characterized by the camera pointing towards the scene, just covering a region of interest (Fig. 1-c) and d)), with the camera motion restricted within a vertical plane.

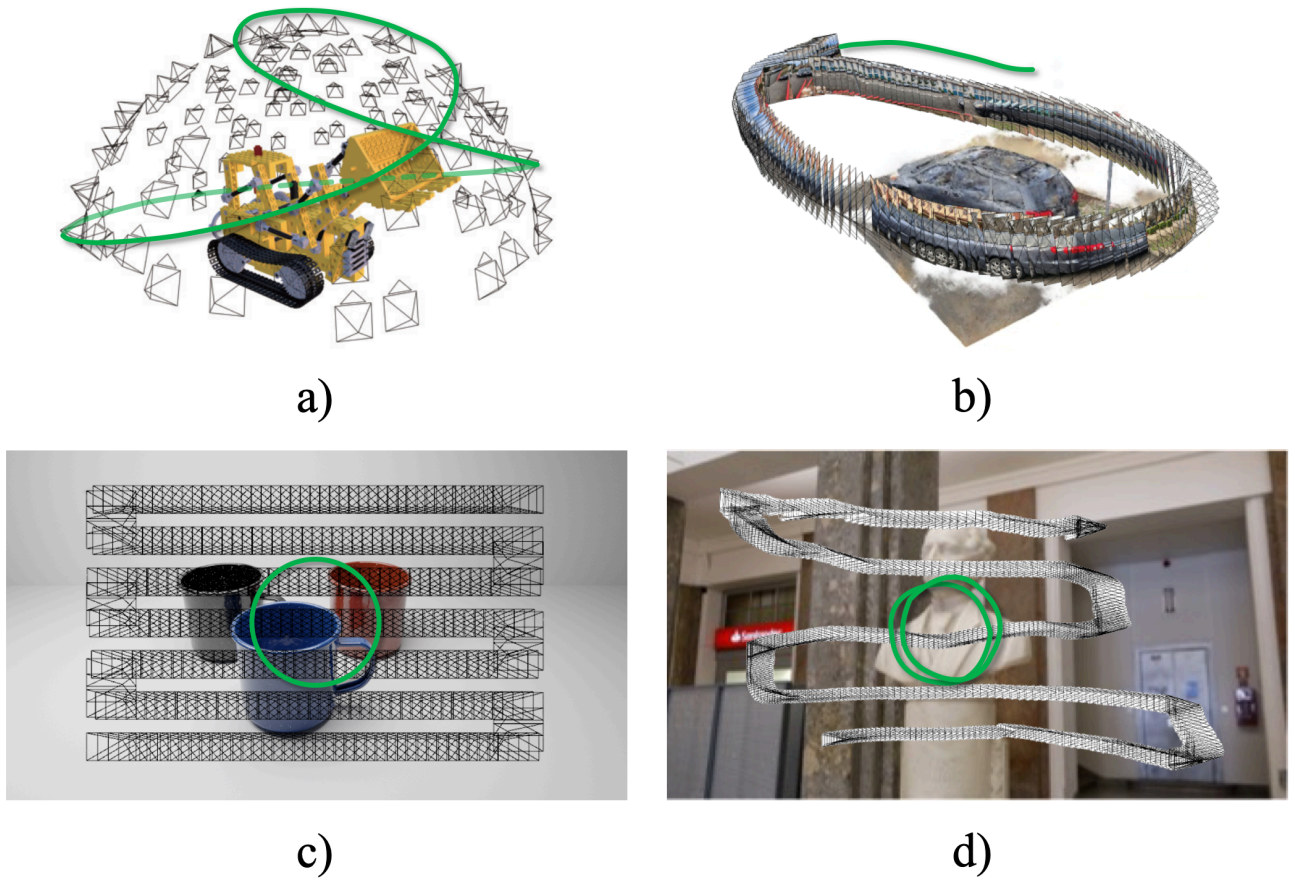

**FIGURE 1. Typical camera paths during training (in grey color) and test (in green color) for a) 360º synthetic [26], b) 360º real [27], c) FF synthetic and d) FF real scenes.**

- **Bounded or unbounded content:** Bounded scenes are composed by a finite number of objects delimited in space, with no background content (e.g., Fig. 1-a)); unbounded scenes correspond to the traditional case, with foreground and background content, as in Fig. 1-d).

### *C. SELECTED NVS METHODS*

From several NVS methods that have been recently proposed, a subset was selected according to the synthesis performance, training and synthesis speed, and suitability to the considered scene classes. Table I presents a summarized description of each selected NVS method and the considered scene classes. It is worthy to note that some combinations of methods and scene classes were excluded from the study, as the resulting synthesis quality was far below acceptable; this was the case for Instant-NGP, Plenoxels, and TensoRF [28] with real scenes. Plenoxels was not applied to FF synthetic scenes due to its design and performance similarities with the DVGO method. Moreover, NeRF++ [24], Mip-NeRF 360 [25], and Nerfacto [29] were specifically designed for unbounded scenes; the first two (NeRF++ and Mip-NeRF 360) for 360º scenes. The remaining methods (Instant-NGP, DVGO [30], Plenoxels, and TensoRF) do not target a specific scene class, but are more suitable for the synthetic scenes synthesis; besides, due to the use of 3D voxel grids in these methods, the training/synthesis times are significantly lower compared to NeRF++ and Mip-NeRF 360.

### *D. ARTIFACTS CHARACTERIZATION*

NVS may result in specific distortions (or artifacts) on the synthesized views, and the characterization of these artifacts may contribute to a deeper understanding of the quality assessment results. Note that such artifacts may not be well modeled in current objective quality metrics. The following artifact types are commonly observed in the synthesized scenes:

- **Floaters:** As the name suggests, floaters correspond to 3D shapes that are not part of the original scene and seem to be suspended in free space (cf. Fig. 2). Floaters are mainly caused by sparse training poses, synthesized poses far away from the training poses, and strong view-dependent effects leading to ambiguities regarding the learned scene opacity values during the NeRF training process [31].
- **Flawed geometry:** This artifact corresponds to the presence of distortions in objects' boundaries, which alter their geometry (cf. Fig. 2). The synthesized object may have flaws in the geometry due to insufficient information about the scene (i.e. the training data is insufficient), or to the shape-radiance ambiguity. The latter concept was introduced in [24], where the authors state that without proper regularization, the NeRF training may learn a wrong scene geometry even when the training images are synthesized with high fidelity (but lacking fidelity for other viewpoints due to ambiguities).
- **Flickering object edges:** This artifact corresponds to color oscillations, along time, on the boundaries of scene objects. Temporal flickering was already present in DIBR-based synthesis, but in the NVS context it is mainly caused by changes in luminosity during the capture of the training images, due to variations in the internal and external camera parameters. As a result, the

**TABLE I. List of selected NVS methods**

| NVS Method | Description | Scenes | | | |
|---|---|---|---|---|---|
| | | (a) | (b) | (c) | (d) |
| **DVGO** [30] | Super-fast convergence approach that adopts a scene representation consisting of a density voxel grid for scene geometry, and a feature voxel grid with a shallow network for complex view-dependent appearance. | ✓ | ✓ | ✓ | ✓ |
| **Instant-NGP** [22] | Achieves fast training by implementing neural graphics primitives with a small neural network, and a multiresolution hash table. | | | ✓ | ✓ |
| **Mip-NeRF 360** [25] | Addresses the synthesis challenges for unbounded 360-degree scenes ; uses a non-linear scene parameterization, online distillation, and a distortion-based regularizer, to overcome blurry renderings. | ✓ | ✓ | ✓ | |
| **NeRF++** [24] | Aims to enhance view synthesis fidelity for unbounded 360-degree scenes by incorporating a novel hierarchical sampling scheme and a shape-prior regularizer. | ✓ | ✓ | ✓ | |
| **Nerfacto** [29] | Combines components from existing NVS methods, seeking a balance between speed and quality. | ✓ | ✓ | ✓ | |
| **Plenoxels** [23] | Uses a sparse voxel grid with spherical harmonics, optimized with a reconstruction loss and a total variation regularizer; comparatively to the seminal NeRF [3], achieves a faster training with a similar view synthesis quality level. | | | | ✓ |
| **TensoRF** [28] | Models and reconstructs the scene radiance field using a 4D tensor representation of a voxel grid, with per-voxel multi-channel features; factorizes the 4D tensor into multiple, low-rank tensor components. | | | ✓ | ✓ |

**(a)** FF real scenes; **(b)** 360º real scenes; **(c)** FF synthetic scenes; **(d)** 360º synthetic scenes

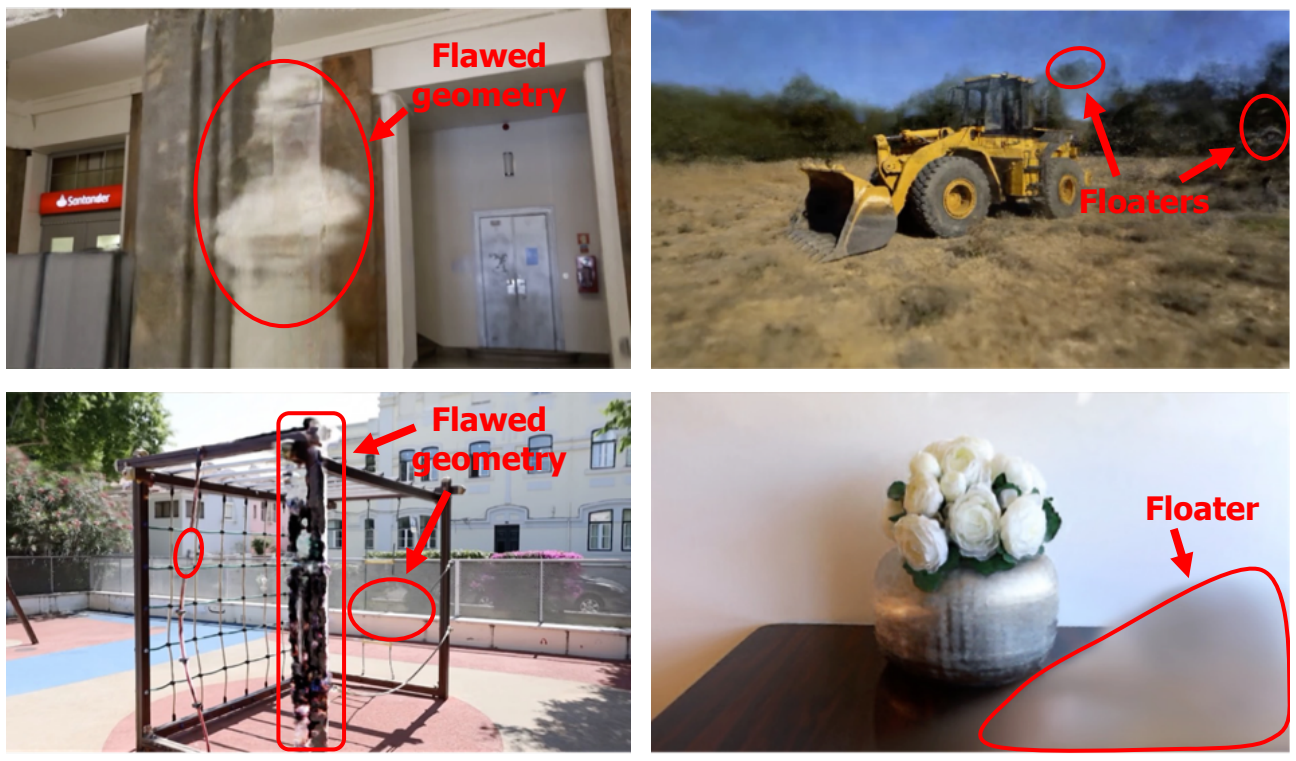

FIGURE 2. Examples of NeRF-related artifacts.

radiance field learning overfits to the observed photometric inconsistencies [32].

## IV. NVS SUBJECTIVE ASSESSMENT AND EVALUATION

This section describes the NeRF-based synthesis process that generates video sequences for the subjective test, as well as the subjective test methodology, the corresponding experimental results, along with some analysis and conclusions.

### *A. NVS CREATION FRAMEWORK*

The NVS framework receives as input a set of training images and a test video. The test video is used to obtain the synthesized video camera trajectory and serves as a reference in the subjective test. Thus, the same camera trajectory is employed for the synthesized and reference video sequences, enabling a double-stimulus evaluation. For real scenes, the training images are selected frames extracted from a video sequence, while for synthetic scenes correspond to different viewing directions. If the scene type is real, pose estimation must be performed; for synthetic scenes, the camera poses are extracted directly from the visual scene. The framework's output is the scores provided by the subjects expressing the synthesized video quality relative to the reference. Fig. 3 shows the NeRF creation and evaluation framework's pipeline which is described as follows:

- **Pose estimation:** This process uses COLMAP structure-from-motion [33], [34] to estimate camera poses. COLMAP was selected due to its popularity and widespread use in the NVS community, having been introduced in the seminal work of [3]. However, this process is not perfect and an error in the estimated camera poses is introduced, typically measured by the mean reprojection error (MRE) metric. The MRE is defined as the mean of the Euclidian distances (measured in pixels) between the reprojections of the 3D points (from the generated COLMAP 3D scene model) into the image plane and their corresponding true projection [35].
- **NeRF training:** This process uses a set of training images, and the camera poses computed in the previous step (or extracted from the synthetic scene) as input. The training involves an optimization procedure that aims to minimize the loss between synthesized and ground truth images, thereby calculating the NVS method parameters. Most of the considered NVS methods use the mean squared error (MSE) between the training and the corresponding synthesized images as loss function (in some cases, additional regularization terms are included). The training configuration best suited to each class of scenes was selected whenever possible. The considered NVS methods were trained with two NVIDIA GeForce RTX 4090 GPUs.

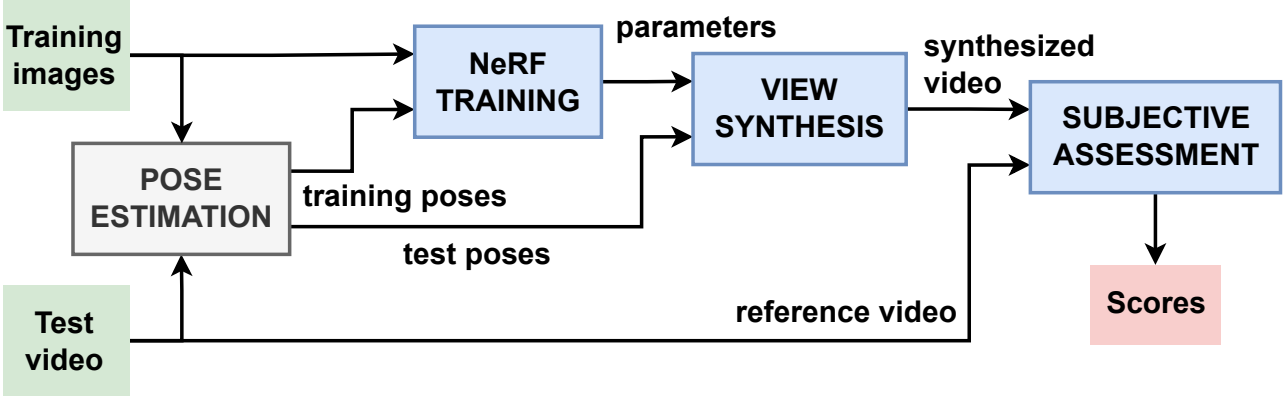

FIGURE 3. NeRF creation and evaluation framework's pipeline.

- **View synthesis:** This process estimates the synthesized image corresponding to a given test pose. For this purpose, a camera ray is computed for each pixel in the image to be synthesized, and a set of 3D points is sampled along this ray. Next, a query is made to the already trained NVS method about the color and opacity values of the sampled 3D points. The synthesized pixel value is then computed based on the accumulation of colors and opacities of the sampled points along the associated camera ray. Finally, a classic volume rendering technique is applied to produce the final output.
- **Subjective assessment:** Subjective assessment is still the most reliable way to evaluate quality. The resulting scores are used as ground-truth to evaluate objective quality metrics. Thus, the synthesized videos quality is assessed by several subjects using a precise and well-defined test methodology. Among those defined in International Standards, the double stimulus continuous quality scale (DSCQS) method was selected [36].

### *B. TEST MATERIAL*

To apply the NVS framework described in the previous section, a set of challenging visual scenes, representative of several NVS use cases, are needed. The followed approach was to select some visual scenes from the popular and well-known Tanks and Temples [37] and Realistic Synthetic 360º [3] datasets, both with 360º captures of real and synthetic scenes, respectively, and capture new FF real and synthetic scenes, that constitute the so-called IST/IT dataset. The new captures are needed since most of the available FF datasets consist of a sparse set of views and thus a reference pristine video cannot be created (i.e. the reference for the double stimulus subjective test). The selected visual scenes were organized into four classes, which are depicted and summarized in Fig. 4, and that can be described as follows:

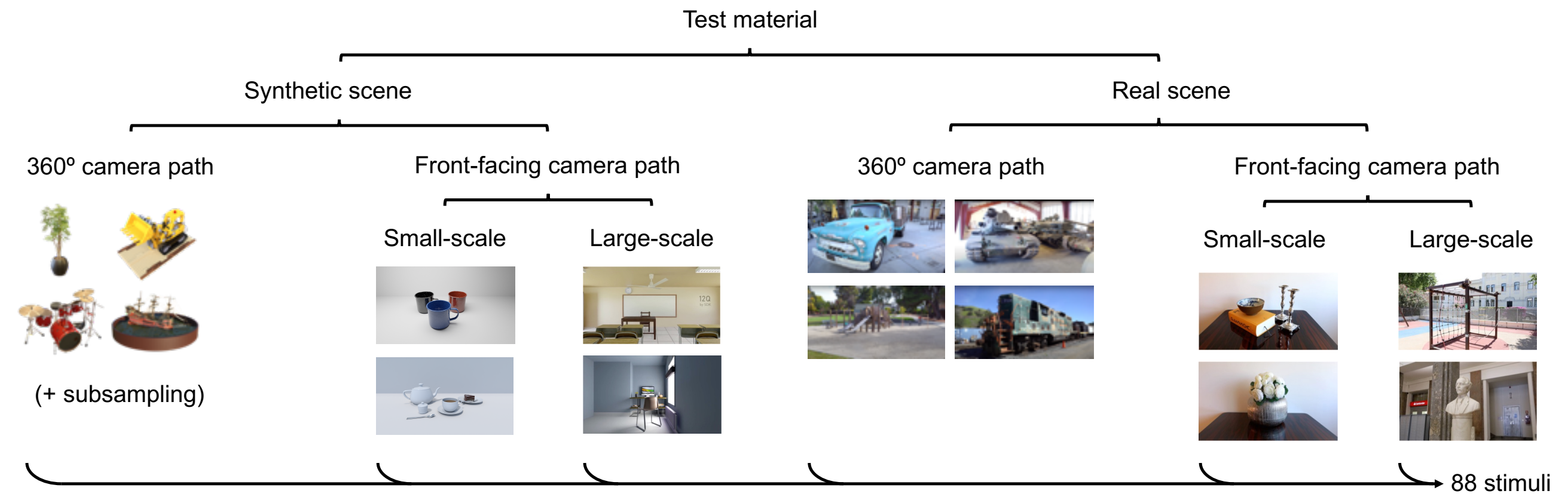

**FIGURE 4. Subjective test material organized by classes of scenes.**

- **360º synthetic scenes:** Due to their popularity, the following visual scenes were selected from [3]: *drums*, *ficus*, *lego*, and *ship*. The *drums* and *ship* scenes are more complex, containing non-Lambertian materials and specular reflection effects. These objects were modeled in Blender [38] and, for each scene, 100 training images and a 8s test video (with 200 frames at 25 fps) were acquired, both with a spatial resolution of 800×800 pixels. The training images follow a semi-spherical inward-facing camera acquisition system, as shown in Section III.B (see also Fig. 1). The test video camera trajectory completes two turns around the semi-sphere while moving up and down along the z-axis on each revolution (similar to Viviani's curve) and with inward-facing camera orientation.
- **FF synthetic scenes:** A new set of visual scenes never used for NeRF evaluation were selected, more precisely four Blender models obtained from [39]. The first two are small-scale scenes, *mugs* and *tea*, and the other two are large-scale scenes, *classroom* and *office*. Small-scale scenes highlight and draw attention to a few objects of interest and are rather different from large-scale scenes due to their simpler objects and backgrounds (such as white walls). For training, the camera path performs a continuous raster scan (from top-right to bottom-left) along some pre-defined vertical plane (see Fig. 1) pointing to the object. Along the camera trajectory, a sparse set of 300 training images were uniformly sampled. The test video has a duration of 10s (with 250 frames at 25 fps) and the camera trajectory follows a circle within the vertical plane used for training (centered on the plane); two loops are completed along the circle during the test video. In both cases (for training and test) the acquisitions have a spatial resolution of 960×540 pixels. The exact camera poses were obtained using the Blender add-on of [40].
- **360º real scenes:** The following visual scenes were selected from [37]: *M60*, *playground*, *train*, and *truck*. The NeRF representation of the visual scene is obtained from a video sequence with 360º camera motion, as described in Section III.B. For each visual scene, the training images were extracted from the video as in [24], resulting in 277, 275, 258, and 226 training images with 1077×546, 1008×548, 982×546, and 980×546 pixels for each scene, respectively. The test videos were also extracted from the video and have a duration of 10s, with 240 and 300 frames at 24 and 30 fps, respectively. Test videos were extracted from a time interval of the captured video different from the one used to select the training images.
- **FF real scenes:** Four FF real scenes were captured with a Canon EOS RP full-frame camera, equipped with a steadicam stabilizer system. Two small-scale visual scenes, *antique* and *flowers*, and two large-scale scenes, *playground2* and *statue*, were acquired. For each visual scene, the aim was to approximately reproduce the camera trajectory previously described for FF synthetic scenes, but the camera motion was always controlled by a human. Thus, it is not a precise camera movement replication of FF synthetic scenes (no gantry is used) but has the advantage of representing a more realistic use case. All the training images (namely 251, 377, 291 and 228 training images, respectively for *antique*, *flowers*, *playground2*, and *statue*) and test videos (10s videos, with 250 frames each) were acquired with a spatial resolution of 1920×1080 pixels.

To display both the synthesized and reference videos side-by-side on a FullHD (1920×1080) display, the test videos of the FF real scenes were spatially down-sampled by a factor of 2 (using bilinear interpolation) and center cropped to 928×522 pixels; for 360º real scenes, videos were only center cropped to 928×522 pixels. For the synthetic scenes, the original resolution of 800×800 and 928×522 pixels (obtained after center cropping) for 360º and FF classes, respectively, were used.

### *C. SUBJECTIVE QUALITY ASSESSMENT METHODOLOGY*

Nowadays, several popular subjective test methodologies are defined in international standards (mostly from ITU). In this work, the synthesized video quality was measured relative to a reference video (same trajectory) using the DSCQS methodology [36]. DSCQS is appropriate for cases where the processed video may have a higher quality than the reference, which can happen in this scenario [13], [15]. At every trial, the subject is shown two videos side-by-side on the display: the reference video and the synthesized video obtained by one of the selected NVS methods (described in Section III.B). The reference and the synthesized video are not labeled, and their location (left or right) is randomly chosen. At every trial, the subject must evaluate both stimuli with a continuous slider with five quality labels (Bad, Poor, Fair, Good, and Excellent) and the selected scores (converted to a value between 0 and 100) are stored.

**TABLE II. Computational efficiency of the selected NVS methods.**

| NVS Method | Tr. time | Syn. time | #Param. |
|---|---|---|---|
| **DVGO** (w/ syn. scenes) | 2.0 mins | 0.36 mins | 4.1 M |
| **DVGO** (w/ real scenes) | 7.7 mins | 1.4 mins | 4.1 M |
| **Instant-NGP** | 4.9 mins | 0.4 mins | 12.6 M |
| **Mip-NeRF 360** | ~ 30 hours | 23.7 mins | 9.9 M |
| **Nerfacto** | 7.2 mins | 1.2 mins | - |
| **NeRF++** | ~ 15 hours | 22.6 mins | 2.4 M |
| **Plenoxels** | 3.5 mins | 0.6 mins | 10 M |
| **TensoRF** | 4.0 mins | 6.4 mins | 27 M |

A total of 88 pairs of stimuli (56 synthesized synthetic scene videos plus 32 synthesized real scene videos, together with the respective reference videos) were evaluated. To avoid fatigue, two separate sessions were performed with at least one day in between. Moreover, to prevent subjects from being influenced by different types of scenes (which often have varying quality ranges and use cases), the first session was dedicated to synthetic scenes, while the second session focused on real scenes. At the beginning of each test session, the scope of the test was explained, followed by a training session that preceded the subjective test. The subjective assessment was carried out on an ASUS ProArt PA32UC-K 4K HDR monitor configured to have a spatial resolution of 1920×1080 pixels. A total of 22 non-expert viewers, 16 male and 6 female, aged between 21 and 57 years, took part in both sessions of the test.

Regarding data processing, difference mean opinion score (DMOS) values were firstly computed for each pair of video sequences according to [41]:

$$\mathrm{DMOS} = \mathrm{MOS}_{\mathrm{syn}} - \mathrm{MOS}_{\mathrm{ref}} + 5 \qquad (1)$$

where $\mathrm{MOS}_{\mathrm{syn}}$ is the MOS value of the synthesized video sequence and $\mathrm{MOS}_{\mathrm{ref}}$ is the MOS value of the respective reference video sequence. To obtain $\mathrm{MOS}_{\mathrm{syn}}$ and $\mathrm{MOS}_{\mathrm{ref}}$, a linear scale conversion from [0, 100] to [1, 5] was first applied, followed by the mean calculation of the subjects' opinion scores for the synthesized and reference stimuli. The DMOS formulation of (1) favors a quality interpretation similar to the MOS values, since a higher DMOS corresponds to higher quality scores and vice versa. According to this formulation, DMOS values greater than 5 indicate that the synthesized video has higher perceived quality than the reference video.

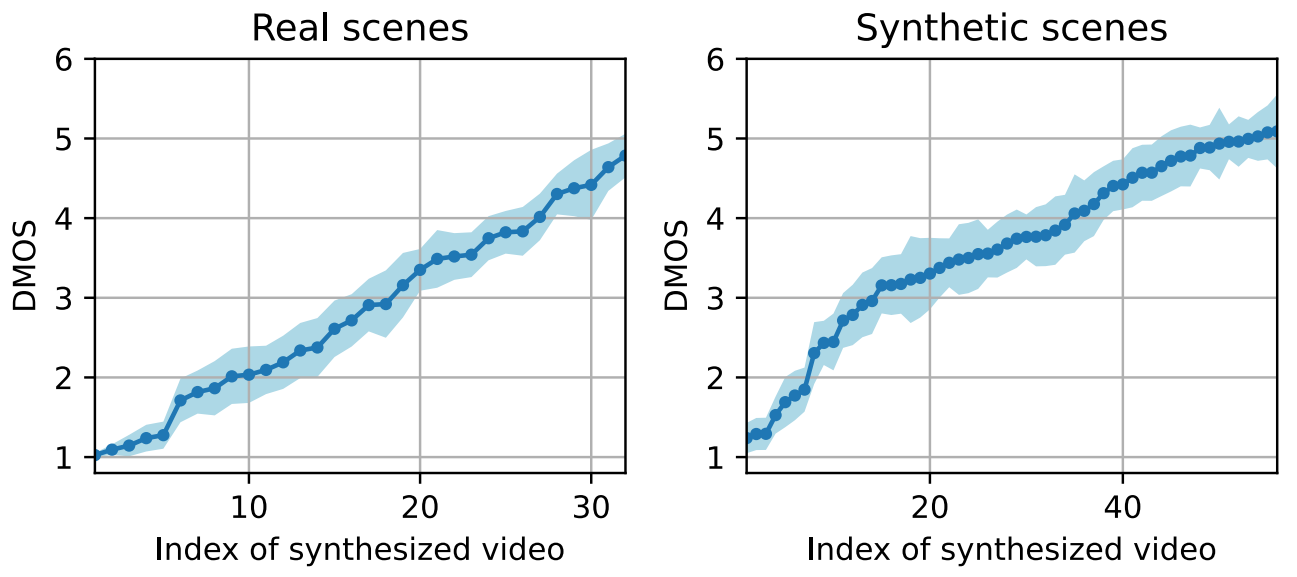

**FIGURE 5. Sorted DMOS values of the synthesized stimuli with the corresponding 95% confidence intervals.**

Finally, outlier detection was performed according to the procedure defined in ITU-R BT 500.15 [36], where the scores of each subject are compared to the overall mean value and standard deviation, to identify large deviations. Consequently, the scores of two subjects were removed for the first session, and the scores of one subject were removed for the second session.

### *D. EXPERIMENT RESULTS AND ANALYSIS*

First, the NVS methods' computational efficiency was characterized by measuring the average training and synthesis times (see Table II). On average, the voxel-based methods (namely DVGO, Instant-NGP, Plenoxels, and TensoRF) have training and synthesis times two orders of magnitude lower than the other methods (except for Nerfacto). On the other hand, voxel-based methods require a much higher memory storage (especially TensoRF).

The obtained DMOS values, and corresponding 95% confidence intervals, are depicted in ascending order in Fig. 5, for all real and synthetic scenes. This figure shows that the quality range (between 1 and 5) is fully covered for both real and synthetic scenes, and that the synthesis of synthetic scenes reached higher DMOS values (even higher than 5) than of real scenes.

For all the previously defined classes in the database, a cumulative distribution function (CDF) was applied to the DMOS data. Fig. 6 shows the results obtained, which represent the percentage of DMOS values (vertical axis) less than or equal to a certain value (horizontal axis). As shown, the 360º real scenes curve is always above the 360º synthetic scenes curve. The same behavior can be observed for FF real scenes when compared to FF synthetic scenes. This is expected since real scenes are harder to synthesize than synthetic scenes, due to the pose estimation error and the presence of complex backgrounds, and thus DMOS scores are more concentrated in the lowest DMOS range.

Another interesting observation from Fig. 6 is that the curve of FF real scenes crosses the 360º real scenes. This can be justified by the fact that some of the NVS methods selected for FF real scenes do not perform very well (low DMOS values) while others achieve very high performance (high DMOS values). For 360º real scenes, the DMOS scores are more concentrated in the mid-range interval of quality (from 2 to 4 of DMOS).

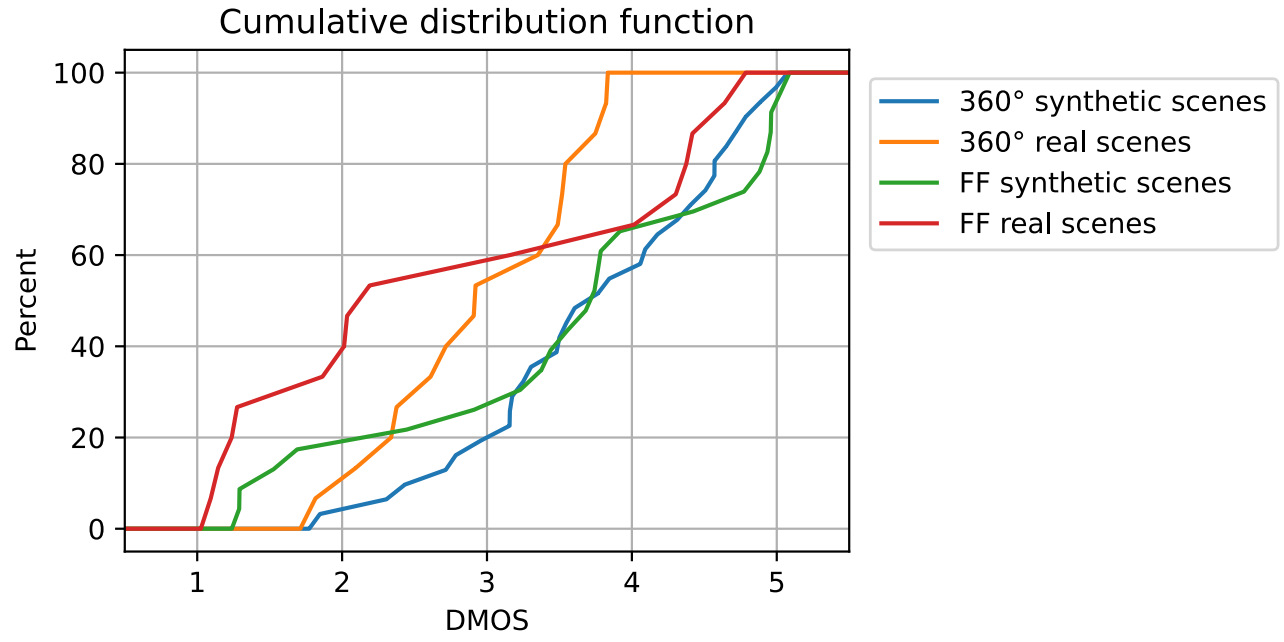

FIGURE 6. CDF of the DMOS for each class of scenes.

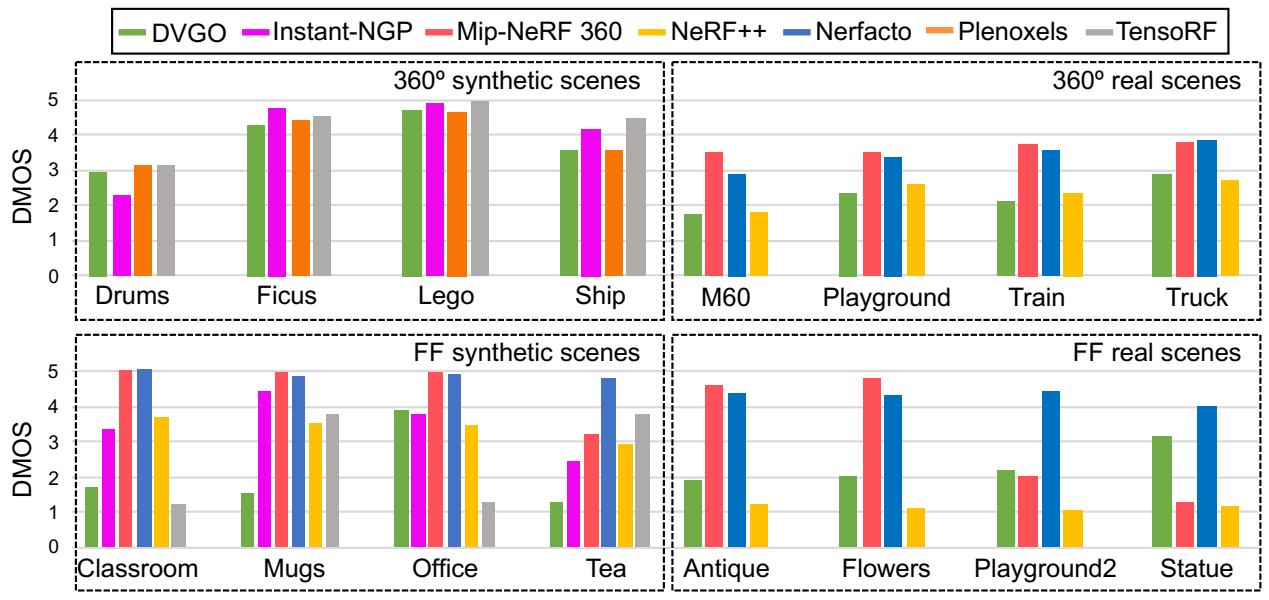

FIGURE 7. DMOS of each synthesized scene, discriminated by the used NVS method.

A more detailed view of the quality scores is presented in Fig. 7, which illustrates the DMOS values obtained for each scene and method, grouped by classes of scenes. Within the same class, there are scenes with varying qualities, revealing that the inherent characteristics of the visual scenes may influence the synthesis performance significantly. The *ficus* and *lego* scenes (which do not contain non-Lambertian materials and specular reflection effects) stand out as the highest evaluated quality scenes, with DMOS values above 4. Furthermore, in FF real scenes, the small-scale scenes (*antique* and *flowers*) obtained around 26% higher qualities, on average, than the large-scale scenes (*playground2* and *statue*). This may explain why the real scenes curves behave differently for higher and lower CDF percentages, as previously observed. The same happens for FF synthetic scenes, where TensoRF presents the worst performance for the large-scale scenes *classroom* and *office* (compared to small-scale scenes *mugs* and *tea*) within the FF synthetic scenes.

Finally, Table III presents for each NVS method the DMOS average for all scenes within each class. The following conclusions can be taken from Table III and Fig. 7:

- **NVS methods:** Nerfacto stands out significantly as the best method (in terms of quality) for synthesizing FF scenes, achieving on average DMOS values of around 4.28 for FF real scenes and almost 5 for FF synthetic scenes. Mip-NeRF 360, which like Nerfacto is also an NN-based method, also has high performance for FF synthetic scenes, but stands out as the best method for 360° real scenes. For 360° synthetic scenes, the voxel-based TensoRF has, on average, the highest DMOS scores (of around 4.31). Considering also the computational performance (see Table II), it can be concluded that the best trade-off between synthesis quality and computational efficiency is achieved by Nerfacto for real and FF synthetic scenes and TensoRF for 360° synthetic scenes.
- **Scene classes:** 360° synthetic scenes are the only class of scenes with an average DMOS rating greater than 4. This is the only class with bounded scenes and NVS methods are easier to train and obtain high quality results for this class, compared to other classes where visual content exists at any direction from the camera, such as FF synthetic scenes. Moreover, the 360° real scenes have, on average, higher DMOS values than FF real scenes. The main reason is that both NeRF++ and Mip-NeRF 360 methods were designed with a ray reparameterization specifically for 360° captures, which does not perform very well for large-scale FF scenes. Actually, the DVGO and Nerfacto methods (and even Mip-NeRF 360 for small scale scenes) have better performance for FF real scenes compared to 360° real scenes.

TABLE III. Average DMOS values for each NVS method discriminated by the visual scene class.

| NVS Method | Real | | Synthetic | |
|---|---|---|---|---|
| | FF | 360° | FF | 360° |
| **DVGO** | 2.31 | 2.28 | 2.11 | 3.90 |
| **Instant-NGP** | N/A | N/A | 3.50 | 4.04 |
| **Mip-NeRF 360** | 3.18 | **3.64** | 4.54 | N/A |
| **NeRF++** | 1.13 | 2.37 | 3.40 | N/A |
| **Nerfacto** | **4.28** | 3.41 | **4.92** | N/A |
| **Plenoxels** | N/A | N/A | N/A | 3.94 |
| **TensoRF** | N/A | N/A | 2.52 | **4.31** |
| **Average** | 2.72 | 2.92 | 3.50 | 4.05 |

## V. OBJECTIVE METRICS PERFORMANCE ASSESSMENT

In this section, a thorough performance assessment of several popular objective quality metrics is conducted. The selected metrics, representative of the current state-of-the-art on visual quality assessment, are first described. Then, a detailed analysis of the metrics performance results is presented, drawing insights about the suitability of the considered metrics for assessing the quality of NVS.

### *A. SELECTED OBJECTIVE ASSESSMENT METRICS*

The objective quality metrics selection was based on several criteria, notably: *i)* metrics that are typically applied in the NVS literature, namely PSNR-Y, PSNR-YUV, SSIM [5], MS-SSIM [6], LPIPS [7], and FovVideoVDP [8], since it is important to know whether they are in fact suitable for assessing the quality of NVS content; *ii)* metrics that have shown good performance in several image and video applications, such as PSNR-HVS [42], IW-SSIM [43], VIF and VIFp [44], FSIM [45], VSI [46], MAD [47], GMSD [48], and NLPD [49]; *iii)* learning-based metrics (besides LPIPS), namely ST-LPIPS [50], DISTS [51], and VMAF [52]; *iv)* MSE, since every selected NVS method includes it in the loss function. All metrics are full-reference.

Most of the selected metrics were designed for the quality assessment of still images (IQA metrics), with FovVideoVDP and VMAF being the exceptions as they were specifically tailored for video (VQA metrics); the latter, being video metrics, have the advantage of accounting for temporal distortions. In this paper, the latter were directly applied to the reference and synthesized videos, whereas the former were applied to individual frames of both videos, followed by the averaging of the frame-based quality scores to obtain a global video score. In terms of the metrics implementation, while for DISTS, FovVideoVDP, IW-SSIM, LPIPS, PSNR-HVS, and VMAF the code was provided by the authors in their respective publications, IQA libraries were used for the remaining metrics. Table IV presents a comprehensive overview of the selected metrics, including their succinct descriptions, reference to the source code, and the used color space. In [53], [54], the authors have suggested that a disparity compensation should be performed between DIBR-based synthesized and reference images, before applying conventional quality metrics (such as SSIM). For NVS, we have noticed that even though the MRE is around 0.86 and 1.04 pixels for 360º real and FF real scenes, respectively, the NeRF pipeline may lead to an increase of the pose errors. Thus, the shift between reference and synthesized images can be significantly higher than the corresponding MRE. Therefore, disparity compensation was applied, using a simple block-matching algorithm with a macroblock size of 32×32 pixels and a search window of 12 pixels (the exceptions to this were *antique* and *flowers*, when synthesized with NeRF++, where a search window of 50 pixels was selected, as disparities of more than 12 pixels were present).

**TABLE IV. List of Selected Objective Quality Metrics.**

| Metric | Color space | Description |
|---|---|---|
| **MSE** | RGB | Average squared difference between corresponding pixels of the two images. |
| **PSNR-Y** | Y | Ratio between the maximum power of the image luminance component, and the power of its distortion, measured by the MSE. |
| **PSNR-YUV** | YUV | PSNR applied on the YUV color space. |
| **PSNR-HVS** [42] | Y | PSNR with the MSE computed in the DCT domain, using frequency dependent weights. |
| **SSIM** [5] | Y | Quantifies the similarity between the two images in terms of luminance, contrast, and structure. |
| **MS-SSIM** [6] | Y | Multi-scale variation of SSIM. |
| **IW-SSIM** [43] | Y | A variation of SSIM using information content weighted pooling. |
| **VIF** [44] | Y | Considers natural scene statistics in the wavelet domain for content fidelity comparison between the two images. |
| **VIFp** [44] | Y | Pixel domain version of VIF. |
| **FSIM** [45] | LAB | Normalized average value of features similarity between the two images. |
| **VSI** [46] | RGB | Uses visual saliency information both as a quality feature and as a weighting function at the pooling stage. |
| **MAD** [47] | Y | Models adaptative strategies of the human visual system, combining a distortion detection strategy and an appearance-based strategy. |
| **LPIPS** [7] | RGB | Measures the perceptual similarity between the two images, based on neural network learned features. |
| **ST-LPIPS** [50] | RGB | A variation of LPIPS tolerant to small pixel disparities among between the two images. |
| **DISTS** [51] | RGB | Structure and texture similarity measurements (SSIM-like) between corresponding feature maps of the two images. |
| **GMSD** [48] | Y | Computes the pixel-wise gradient magnitude similarity (GMS) between two images, followed by pooling based on the standard deviation of the GMS map. |
| **NLPD** [49] | Y | Mimics the nonlinear transformations of the early visual system, as local luminance subtraction and local gain control, and combines these values using weighted lp-norms. |
| **VMAF** [52] | YUV | Merges existing metrics and image feature components using a support vector machine. |
| **FovVideoVDP** [8] | RGB | Considers spatial, temporal, and peripherical aspects related with the foveate vision effect. |

### *B. METRICS PERFORMANCE RESULTS AND ANALYSIS*

To evaluate the metrics' performance, the Pearson Linear Correlation Coefficient (PLCC), the Spearman Rank Order Correlation Coefficient (SROCC), and the Perceptually Weighted Rank Correlation (PWRC) were calculated between the predicted (by the metric) and reference (from the subjective tests) DMOS values. Unlike traditional rank correlation measures, as SROCC, the PWRC measure emphasizes the ranking accuracy of high-quality images while considering subjective uncertainty between images of similar quality [55]. It introduces the concept of a sensory threshold (ST), which defines the point at which differences between images become perceptible to human viewers. The PWRC method provides a sorting accuracy-sensory threshold (SA-ST) curve that shows how well a metric aligns with human perception across different levels of visual sensitivity. Moreover, given that subjective quality scores frequently saturate at the extremes of rating scales, it is recommended to employ a non-linear mapping between objective metrics and DMOS values [56]; in this paper, such mapping has been implemented using cubic regression, as defined in ITU-T P.1401 [56]. The statistical significance analysis of the quality metrics is presented in Appendix A. The results obtained support the subsequent analysis of the performance of the selected objective metrics.

The PLCC and SROCC coefficients, as well as the PWRC values for a null sensory threshold, are presented in Tables V, VI, and VII, respectively. In all tables, a color scale varying from red to green was used to represent the coefficients variation from worst to best performance, respectively. Regarding the best quality metrics, according to the scene class, it can be figured out that: *i)* for synthetic scenes, MSE-RGB and PSNR-YUV are the top performers, achieving PLCC, SROCC, and PWRC values of 0.87, 0.83, and 0.73, respectively; *ii)* for real scenes, ST-LPIPS and DISTS lead with PLCC, SROCC, and PWRC values of 0.68, 0.64, and 0.56; *iii)* for the real scenes with disparity compensation, VMAF and VSI excel with PLCC, SROCC, and PWRC of 0.80, 0.77, and 0.68, respectively. When all scene classes are considered, DISTS and FovVideoVDP (without disparity compensation) and IW-SSIM (with disparity compensation)

**TABLE V. PLCC objective metrics' performance results.**

| Metric | Synthetic 360 | Synthetic FF | Synthetic All | Real 360 | Real FF | Real All | All | Real* 360* | Real* FF* | Real* All* | All* |
|---|---|---|---|---|---|---|---|---|---|---|---|
| MSE-RGB | 0.94 | 0.89 | 0.87 | 0.59 | 0.78 | 0.57 | 0.65 | 0.70 | 0.66 | 0.59 | 0.73 |
| PSNR-Y | 0.94 | 0.84 | 0.86 | 0.57 | 0.66 | 0.51 | 0.67 | 0.65 | 0.70 | 0.56 | 0.70 |
| PSNR-YUV | 0.94 | 0.85 | 0.86 | 0.63 | 0.63 | 0.49 | 0.64 | 0.76 | 0.64 | 0.52 | 0.67 |
| PSNR-HVS | 0.93 | 0.85 | 0.86 | 0.58 | 0.69 | 0.53 | 0.69 | 0.60 | 0.81 | 0.64 | 0.73 |
| SSIM | 0.90 | 0.80 | 0.69 | 0.70 | 0.70 | 0.47 | 0.60 | 0.82 | 0.72 | 0.57 | 0.62 |
| MS-SSIM | 0.95 | 0.88 | 0.77 | 0.71 | 0.72 | 0.53 | 0.65 | 0.78 | 0.80 | 0.63 | 0.69 |
| IW-SSIM | 0.93 | 0.83 | 0.80 | 0.66 | 0.67 | 0.53 | 0.68 | 0.82 | 0.87 | 0.77 | 0.78 |
| VIF | 0.91 | 0.74 | 0.77 | 0.67 | 0.61 | 0.51 | 0.66 | 0.77 | 0.80 | 0.73 | 0.75 |
| VIFp | 0.90 | 0.75 | 0.75 | 0.77 | 0.63 | 0.51 | 0.63 | 0.86 | 0.78 | 0.70 | 0.71 |
| FSIM | 0.93 | 0.87 | 0.73 | 0.69 | 0.72 | 0.48 | 0.66 | 0.83 | 0.81 | 0.65 | 0.63 |
| VSI | 0.93 | 0.77 | 0.78 | 0.65 | 0.75 | 0.60 | 0.68 | 0.77 | 0.83 | 0.78 | 0.77 |
| MAD | 0.87 | 0.63 | 0.71 | 0.76 | 0.67 | 0.56 | 0.67 | 0.68 | 0.78 | 0.67 | 0.71 |
| LPIPS | 0.87 | 0.90 | 0.55 | 0.75 | 0.75 | 0.67 | 0.60 | 0.90 | 0.87 | 0.76 | 0.64 |
| ST-LPIPS | 0.92 | 0.92 | 0.77 | 0.88 | 0.82 | 0.68 | 0.66 | 0.87 | 0.90 | 0.71 | 0.68 |
| DISTS | 0.94 | 0.83 | 0.81 | 0.61 | 0.71 | 0.67 | 0.76 | 0.70 | 0.78 | 0.72 | 0.76 |
| GMSD | 0.95 | 0.81 | 0.82 | 0.74 | 0.74 | 0.52 | 0.67 | 0.84 | 0.81 | 0.68 | 0.72 |
| NLPD | 0.95 | 0.87 | 0.86 | 0.65 | 0.73 | 0.52 | 0.69 | 0.70 | 0.79 | 0.61 | 0.72 |
| VMAF | 0.87 | 0.76 | 0.79 | 0.81 | 0.64 | 0.61 | 0.64 | 0.78 | 0.92 | 0.80 | 0.77 |
| FovVideoVDP | 0.96 | 0.90 | 0.80 | 0.49 | 0.76 | 0.65 | 0.70 | 0.41 | 0.83 | 0.65 | 0.73 |

*After disparity compensation

**TABLE VI. SROCC objective metrics' performance results.**

| Metric | Synthetic 360 | Synthetic FF | Synthetic All | Real 360 | Real FF | Real All | All | Real* 360* | Real* FF* | Real* All* | All* |
|---|---|---|---|---|---|---|---|---|---|---|---|
| MSE-RGB | 0.95 | 0.72 | 0.82 | 0.57 | 0.64 | 0.51 | 0.69 | 0.49 | 0.60 | 0.54 | 0.71 |
| PSNR-Y | 0.95 | 0.77 | 0.82 | 0.57 | 0.64 | 0.50 | 0.68 | 0.50 | 0.64 | 0.48 | 0.70 |
| PSNR-YUV | 0.95 | 0.74 | 0.83 | 0.59 | 0.61 | 0.48 | 0.66 | 0.54 | 0.55 | 0.46 | 0.68 |
| PSNR-HVS | 0.95 | 0.77 | 0.81 | 0.57 | 0.66 | 0.48 | 0.69 | 0.51 | 0.70 | 0.50 | 0.73 |
| SSIM | 0.87 | 0.67 | 0.65 | 0.64 | 0.52 | 0.31 | 0.59 | 0.79 | 0.64 | 0.49 | 0.62 |
| MS-SSIM | 0.96 | 0.68 | 0.74 | 0.74 | 0.66 | 0.44 | 0.67 | 0.74 | 0.70 | 0.51 | 0.70 |
| IW-SSIM | 0.93 | 0.68 | 0.74 | 0.68 | 0.67 | 0.48 | 0.68 | 0.81 | 0.87 | 0.73 | 0.75 |
| VIF | 0.91 | 0.64 | 0.72 | 0.67 | 0.61 | 0.43 | 0.65 | 0.77 | 0.78 | 0.65 | 0.71 |
| VIFp | 0.92 | 0.67 | 0.74 | 0.81 | 0.63 | 0.45 | 0.65 | 0.89 | 0.77 | 0.68 | 0.70 |
| FSIM | 0.92 | 0.67 | 0.62 | 0.77 | 0.58 | 0.39 | 0.64 | 0.81 | 0.80 | 0.63 | 0.61 |
| VSI | 0.94 | 0.58 | 0.71 | 0.69 | 0.62 | 0.56 | 0.67 | 0.71 | 0.81 | 0.77 | 0.73 |
| MAD | 0.83 | 0.58 | 0.69 | 0.71 | 0.64 | 0.47 | 0.67 | 0.54 | 0.77 | 0.62 | 0.70 |
| LPIPS | 0.74 | 0.86 | 0.39 | 0.75 | 0.78 | 0.59 | 0.55 | 0.89 | 0.87 | 0.72 | 0.57 |
| ST-LPIPS | 0.90 | 0.87 | 0.77 | 0.86 | 0.80 | 0.61 | 0.70 | 0.87 | 0.85 | 0.65 | 0.71 |
| DISTS | 0.94 | 0.69 | 0.77 | 0.64 | 0.70 | 0.64 | 0.72 | 0.70 | 0.81 | 0.70 | 0.72 |
| GMSD | 0.95 | 0.67 | 0.75 | 0.80 | 0.54 | 0.41 | 0.67 | 0.77 | 0.74 | 0.61 | 0.72 |
| NLPD | 0.95 | 0.70 | 0.81 | 0.61 | 0.66 | 0.43 | 0.68 | 0.71 | 0.70 | 0.53 | 0.71 |
| VMAF | 0.89 | 0.57 | 0.74 | 0.75 | 0.76 | 0.50 | 0.67 | 0.61 | 0.88 | 0.70 | 0.73 |
| FovVideoVDP | 0.96 | 0.70 | 0.75 | 0.45 | 0.71 | 0.44 | 0.68 | 0.39 | 0.81 | 0.51 | 0.71 |

*After disparity compensation

**TABLE VII. PWRC objective metrics' performance results.**

| Metric | Synthetic 360 | Synthetic FF | Synthetic All | Real 360 | Real FF | Real All | All | Real* 360* | Real* FF* | Real* All* | All* |
|---|---|---|---|---|---|---|---|---|---|---|---|
| MSE-RGB | 0.82 | 0.56 | 0.66 | 0.43 | 0.45 | 0.33 | 0.49 | 0.33 | 0.42 | 0.34 | 0.51 |
| PSNR-Y | 0.81 | 0.63 | 0.72 | 0.43 | 0.46 | 0.31 | 0.48 | 0.36 | 0.46 | 0.27 | 0.50 |
| PSNR-YUV | 0.81 | 0.60 | 0.73 | 0.51 | 0.43 | 0.29 | 0.46 | 0.37 | 0.38 | 0.34 | 0.48 |
| PSNR-HVS | 0.80 | 0.62 | 0.72 | 0.43 | 0.51 | 0.45 | 0.50 | 0.37 | 0.54 | 0.30 | 0.55 |
| SSIM | 0.72 | 0.51 | 0.47 | 0.46 | 0.39 | 0.12 | 0.39 | 0.68 | 0.48 | 0.29 | 0.43 |
| MS-SSIM | 0.84 | 0.54 | 0.56 | 0.56 | 0.50 | 0.25 | 0.48 | 0.64 | 0.56 | 0.31 | 0.51 |
| IW-SSIM | 0.78 | 0.53 | 0.56 | 0.50 | 0.36 | 0.30 | 0.49 | 0.59 | 0.73 | 0.55 | 0.57 |
| VIF | 0.76 | 0.48 | 0.56 | 0.49 | 0.25 | 0.25 | 0.47 | 0.63 | 0.61 | 0.45 | 0.53 |
| VIFp | 0.77 | 0.51 | 0.55 | 0.64 | 0.46 | 0.28 | 0.41 | 0.76 | 0.62 | 0.50 | 0.52 |
| FSIM | 0.79 | 0.53 | 0.49 | 0.57 | 0.17 | 0.20 | 0.45 | 0.67 | 0.66 | 0.41 | 0.42 |
| VSI | 0.79 | 0.42 | 0.54 | 0.51 | 0.52 | 0.42 | 0.48 | 0.56 | 0.66 | 0.59 | 0.55 |
| MAD | 0.66 | 0.40 | 0.51 | 0.53 | 0.47 | 0.27 | 0.48 | 0.43 | 0.61 | 0.43 | 0.51 |
| LPIPS | 0.60 | 0.73 | 0.27 | 0.53 | 0.62 | 0.38 | 0.38 | 0.73 | 0.73 | 0.52 | 0.40 |
| ST-LPIPS | 0.75 | 0.73 | 0.57 | 0.68 | 0.62 | 0.40 | 0.50 | 0.67 | 0.68 | 0.43 | 0.51 |
| DISTS | 0.80 | 0.58 | 0.61 | 0.44 | 0.70 | 0.56 | 0.54 | 0.55 | 0.66 | 0.53 | 0.55 |
| GMSD | 0.82 | 0.52 | 0.60 | 0.62 | 0.35 | 0.22 | 0.45 | 0.62 | 0.60 | 0.42 | 0.54 |
| NLPD | 0.81 | 0.57 | 0.65 | 0.43 | 0.35 | 0.24 | 0.47 | 0.58 | 0.54 | 0.34 | 0.52 |
| VMAF | 0.71 | 0.43 | 0.57 | 0.59 | 0.68 | 0.61 | 0.49 | 0.45 | 0.73 | 0.68 | 0.55 |
| FovVideoVDP | 0.84 | 0.58 | 0.58 | 0.31 | 0.56 | 0.28 | 0.56 | 0.30 | 0.66 | 0.32 | 0.52 |

*After disparity compensation

achieve the highest PLCC, SROCC, and PWRC values. The following conclusions can also be taken:

1) **Impact of the metric type:** The PLCC, SROCC, and PWRC scores, together with significance tests, show that the learning-based metrics (namely DISTS, LPIPS, ST-LPIPS, and VMAF) stand out: DISTS is among the best metrics in the overall case, for real scenes (without disparity compensation) and for synthetic scenes, while LPIPS, ST-LPIPS, and VMAF excels in FF and 360º real scenes, with and without disparity compensation. These findings suggest that learning-based metrics can represent content with a rich set of features and obtain a reliable quality score (usually via regression). This is rather notable since the training set used to create the model of these metrics does not include many of the artifacts that are typical of NVS. The DISTS performance highlights its tolerance to geometric distortions which is prevalent in novel view synthesis, and thus is very suitable for NVS quality assessment. Furthermore, VQA metrics also show noteworthy performances, notably FovVideoVDP for the overall case (with and without disparity compensation) and synthetic scenes (being the best metric for 360º synthetic scenes), and VMAF for the aforementioned cases. These results suggest that the temporal dimension of the video contains important information for the quality assessment. Finally, metrics based on pixel-wise differences, such as MSE-RGB, PSNR-Y, PSNR-YUV, and PSNR-HVS, are the most effective for synthetic scenes.
2) **Impact of the camera path:** Tables V, VI, and VII show that for synthetic scenes, the average PLCC, SROCC, and PWRC are, respectively, 12%, 32%, and 40% higher for the 360º camera path than for the FF camera path. This can be justified by the fact that the considered 360º scenes are bounded (while FF scenes are unbounded), resulting in an improved convergence during NeRF training and leading to higher quality syntheses, more easily predicted by the objective metrics. Given that both 360º and FF real scenes are unbounded, there is not a significant difference between their respective PLCC, SROCC, and PWRC values. Actually, FF real scenes tend to reach slightly higher performances than 360º real scenes, without surpassing the 10% difference for PLCC, SROCC, and PWRC, with and without disparity compensation.
3) **Impact of the scene type:** According to Tables V, VI, and VII, the average PLCC, SROCC, and PWRC values for synthetic scenes are, respectively, 40%, 52%, and 77% higher than for real scenes (without disparity compensation); when using disparity compensation, the PLCC, SROCC, and PWRC averages variations decrease to 17%, 21%, and 36%. The performance difference between scene types may be justified by the error associated with the pose estimation process performed by real scenes (not done for synthetic scenes); in particular, the 360º real scene class has an MRE of around 0.86 pixels, while the FF real scene class has an MRE of around 1.04 pixels. The MRE has an impact on the NeRF training and synthesis processes, given that both require the pose information of the training and synthesized images. Both processes lead to a pose error amplification, that causes high spatial disparities between synthesis and reference images (between 1 and 45 pixels of observed disparity). Therefore, metrics based on the difference between corresponding pixels have a higher performance for synthetic than for real scenes. Moreover, the new types of artifacts (namely the so-called floaters), particularly conspicuous in real scenes due to their inherent complexity, may also contribute to the difference on the PLCC, SROCC, and PWRC between real and synthetic scenes.

4) **Impact of the disparity compensation:** Tables V, VI, and VII show that the metrics' performance benefits greatly from disparity compensation, especially within real scenes. For 360º real scenes, VIFp and LPIPS achieved PLCC, SROCC, and PWRC values between 0.73 and 0.90, respectively, and for FF real scenes, IW-SSIM, LPIPS, and VMAF obtained PLCC, SROCC, and PWRC between 0.73 and 0.92, respectively. For both real scenes and the overall case, IW-SSIM now outperforms DISTS and FovVideoVDP, which had previously the leading performance. ST-LPIPS (robust to small disparities) has the best performance within real scenes without disparity compensation. However, LPIPS with disparity compensation can handle higher disparities better than ST-LPIPS, showing the importance of the disparity compensation.
5) **Statistical significance analysis of objective assessment:** In addition to evaluating the PLCC, SROCC, and PWRC, the performance differences between objective metrics were assessed for statistical significance using the procedure suggested in [57]. The statistical significance analysis and results are presented in Appendix A. These results show that DISTS is the only metric that is consistently present in the best significance group for every scene class under analysis. When there is no disparity compensation, both PSNR-HVS and NLPD are on the best significance groups. With disparity compensation, VMAF, VSI, IW-SSIM, and GMSD are included in the best significance groups. The statistical significance results confirm the validity of the conclusions drawn before, in Section V.

## VI. CONCLUSIONS

This paper addresses the problem of quality assessment of NVS methods. To tackle this challenge, a subjective assessment test was conducted for several popular NVS methods, along with a characterization of the performance for several well-known IQA and VQA objective quality metrics.

The subjective assessment results show that most NVS methods still struggle with real scenes, especially outdoor scenes with complex backgrounds. More importantly, there is not a single method that is consistently the best for all visual scene classes. Nerfacto, Mip-NeRF 360 and TensoRF reach the highest subjective scores for specific classes of visual content.

Regarding the objective quality metrics performance characterization, the 19 metrics considered underperform for real scenes while obtain satisfactory results for synthetic scenes (especially those without background). One of the main reasons for the poor performance is the large geometric deformations (mostly translation) that occur in real scenes, due to errors in the pose estimation process that is needed for both NeRF training and synthesis. The DISTS quality metric has the highest performance across all scene classes; however, when disparity compensation is performed between the synthesis and reference video sequences, IW-SSIM turns out to be the best metric. For synthetic scenes, pose estimation is not necessary and MSE-based metrics perform reasonably well. For real scenes, all the considered learning-based objective metrics are on the best group of metrics.

## VII. NVS QUALITY ASSESSMENT: CHALLENGES AND WAYS FORWARD

The experimental results presented in this work have led to several important conclusions in the field of NVS quality assessment. In connection with these conclusions, this section highlights challenges and proposes potential ways to advance this field:

- **NVS methods complexity-quality tradeoff:** The ongoing development of NVS methods presents significant challenges in balancing synthesis quality and computational efficiency. This study highlights Nerfacto as the best compromise between these two aspects, but new approaches like 3D Gaussian Splatting (3DGS) [58] represent nowadays a promising paradigm with strong potential in both quality and efficiency. The subjective assessment of 3DGS-based methods remains underexplored, with a critical need for further research.
- **NVS methods model optimization:** In any case, both NeRF and 3DGS use objective quality metrics (like MSE, PSNR, and SSIM) during training primarily for measuring the quality of scene reconstruction and rendering. These metrics are essential to guide the learning process by providing feedback on how well the model is capturing the scene. This study helps to identify the most promising quality metrics that can be used during model optimization to enhance the final synthesized output quality.
- **NVS subjective quality assessment:** The study highlighted in this work underscores the significance of using subjective test methodologies that consider the possibility for distorted sequences to receive higher scores than their reference sequence. In the NVS domain, reference sequences do not always guarantee superior perceived quality, as neural networks can enhance the scene's inherent features. The use of the DSCQS methodology, coupled with a DMOS scoring system, proved crucial in uncovering these subtleties. Moreover, gathering more subjective evaluations from a broader range of users can offer deeper insights into user perceptions of NVS content and improve the training and validation of quality assessment metrics for NVS.

- **NVS objective metrics:** A fundamental challenge is that existing 2D quality metrics are not optimized for the specific requirements of NVS. Traditional 2D metrics may fail to capture critical factors in 3D scene reconstruction, such as depth perception, multi-view consistency, and occlusions. Additionally, assessing the quality of NVS is particularly complex due to the variability in synthesis quality across different scene classes. The existing NVS methods can further introduce different types of distortions in the generated images. The performance characterization of the existing objective quality metrics shows that further research is needed to design a metric with high performance for NVS. Experimental results indicate that best practices for developing a new quality metric should include considering the temporal dimension of NVS content, compensating for spatial disparities between reference and synthesized sequences, and adopting a learning-based approach that models the perceptual impact of NVS artifacts, such as floaters.

## APPENDIX A
## STATISTICAL SIGNIFICANCE TEST FOR OBJECTIVE QUALITY METRICS

This appendix complements Section V.B by presenting the results of the statistical significance tests applied to the objective metrics quality predictions. It provides evidence on whether the observed differences in the performance of the selected metrics are statistically significant or merely due to random variation [57]. A statistical t-test was used for this analysis. Considering this objective, the following steps were taken:

1) **Prediction residuals computation:** The first step in the statistical analysis involves computing the prediction residuals, defined as the absolute difference between the predicted DMOS (DMOSp), obtained from each objective quality metric, and the ground-truth subjective quality scores, DMOS. To obtain DMOSp, a non-linear mapping was applied for each objective quality metric, normalizing their outputs to the DMOS scale. A cubic regression was used for the non-linear mapping, as described in Section V of the manuscript. Consequently, a prediction residual (vector of absolute differences) was obtained for each objective quality metric under consideration.
2) **Verification of normality and homogeneity of variance assumptions:** The prediction residuals of each metric were tested for normality and homogeneity of variance, which are critical assumptions for the statistical t-test. Normality was assessed using the Shapiro-Wilk test (when the sample size was less than 50) and the Kolmogorov-Smirnov test (other sample sizes), with a significance level of 0.05. Homogeneity of variance was evaluated using the Levene's test, also with a significance level of 0.05. These tests confirmed that all prediction residuals follow a normal distribution and have equal variances across samples.
3) **Statistical t-test computation:** A two-sided independent t-test (also known as two-sample F-test) was used as the significance test. This test is applied to a pair of the prediction residuals (thus compares two objective quality metrics) that were obtained in step 1 and defines two hypotheses. The null hypothesis assumes that there is no difference between the means of the two prediction residuals, for the pair of metrics under test. The alternative hypothesis states that there is a difference between these means. The following steps are applied:
   - **Optimal significance levels computation:** a popular decision-theoretic approach [59] was employed to define the significance level thresholds. A threshold is defined according to the number of elements and the Cohen's d parameter (difference between the two means divided by the standard deviation) of the pair of prediction residuals. The threshold is defined independently for each pair of prediction residuals.
   - **p-value computation:** If the p-value from the t-test between the two metrics is lower than the corresponding significance level threshold, established in the previous step, the null hypothesis is not rejected, indicating that the quality metrics are statistically the same. Otherwise, the null hypothesis is rejected, indicating that the quality metrics are statistically different from each other.

Table VIII presents the results of the significance tests for all combinations of two quality metrics applied to the context of synthetic scenes, real scenes, real scenes with disparity compensation, all scenes, and all scenes with disparity compensation. The values shown can be of three types, "0", "-", or "1", indicating whether the metric in the row is statistically better (lower prediction residual mean), indistinguishable, or worse, than the metric in the column, respectively. Table IX shows the statistical significance groups of metrics derived from Table VIII, ordered by increasing mean prediction residual values. Metrics within the same group are statistically identical.

**TABLE VIII. Statistical significance test results for all objective quality assessment metrics. each entry is the codeword that represents the test outcome for synthetic scenes, real scenes, real scenes with disparity compensation, all scenes, and all scenes with disparity compensation.**

| | MSE-RGB | PSNR-Y | PSNR-YUV | PSNR-HVS | SSIM | MS-SSIM | IW-SSIM | VIF | VIFp | FSIM | VSI | MAD | LPIPS | ST-LPIPS | DISTS | GMSD | NLPD | VMAF | FVVDP |
|---|---|---|---|---|---|---|---|---|---|---|---|---|---|---|---|---|---|---|---|
| **MSE-RGB** | | ----- | ----- | ----- | 0---0 | 0---0 | 0-1-- | ----- | 0---- | 0---0 | --1-- | 0---- | 0---0 | 0---0 | ---1- | ----- | ----- | --1-- | 0---- |
| **PSNR-Y** | ----- | | ----- | ----- | 0--00 | 0---- | 0-1-- | ----- | 0---- | 0---- | --1-- | 0---- | 0-1-0 | 01--- | -111- | ----- | ----- | --1-- | ----- |
| **PSNR-YUV** | ----- | ----- | | ----- | 0---- | 0---- | 0-1-1 | --1-- | 0-1-- | 0---- | --1-1 | 0---- | 011-- | 011-- | -111- | ----- | ----- | --1-1 | 0---- |
| **PSNR-HVS** | ----- | ----- | ----- | | 0--00 | 0---- | --1-- | ----- | 0---- | 0---0 | --1-- | 0---- | 0-1-0 | 0--00 | ----- | ----- | ----- | --1-- | ----- |
| **SSIM** | 1---1 | 1--11 | 1---- | 1--11 | | ----- | --1-1 | ----1 | ----1 | ----- | 1-1-1 | ----1 | 0111- | -1--- | 11-11 | 1---1 | 1--11 | --1-1 | ---11 |
| **MS-SSIM** | 1---1 | 1---- | 1---- | 1---- | ----- | | ----1 | ----1 | ----- | ----- | --1-1 | ----- | 0---- | ----- | 1--11 | 1---- | 1---- | --1-1 | ----- |
| **IW-SSIM** | 1-0-- | 1-0-- | 1-0-0 | --0-- | --0-0 | ----0 | | ----- | ----0 | ----0 | ----- | ----- | 0---0 | ----0 | ---1- | ----- | --0-- | ----- | --0-- |
| **VIF** | ----- | ----- | --0-- | ----- | ----0 | ----0 | ----- | | ----- | ----0 | ----- | ----- | 0---0 | ----0 | ---1- | ----- | ----- | ----- | ----- |
| **VIFp** | 1---- | 1---- | 1-0-- | 1---- | ----0 | ----- | ----1 | ----- | | ----- | ----- | ----- | 01--- | -1--- | 11-1- | 1---- | 1---- | ----- | ----- |
| **FSIM** | 1---1 | 1---- | 1---- | 1---1 | ----- | ----- | ----1 | ----1 | ----- | | ----1 | ----1 | 01--- | -1--- | 11-11 | 1---1 | 1---1 | --1-1 | ----1 |
| **VSI** | --0-- | --0-- | --0-0 | --0-- | 0-0-0 | --0-0 | ----- | ----- | ----- | ----0 | | ----- | 0---0 | ----0 | ---1- | ----- | --0-- | ----- | --0-- |
| **MAD** | 1---- | 1---- | 1---- | 1---- | ----0 | ----- | ----- | ----- | ----- | ----0 | ----- | | 0---0 | ----0 | 1--1- | 1---- | 1---- | --1-- | ----- |
| **LPIPS** | 1---1 | 1-0-1 | 100-- | 1-0-1 | 1000- | 1---- | 1---1 | 1---1 | 10--- | 10--- | 1---1 | 1---1 | | 1--0- | 1---1 | 1---1 | 1-0-1 | 1---1 | 1-0-1 |
| **ST-LPIPS** | 1---1 | 10--- | 100-- | 1--11 | -0--- | ----- | ----1 | ----1 | -0--- | -0--- | ----1 | ----1 | 0--1- | | ---11 | ----1 | ---11 | ----1 | ---11 |
| **DISTS** | ---0- | -000- | -000- | ----- | 00-00 | 0--00 | ---0- | ---0- | 00-0- | 00-00 | ---0- | 0--0- | 0---0 | ---00 | | ---0- | ----- | ---0- | ----- |
| **GMSD** | ----- | ----- | ----- | ----- | 0---0 | 0---- | ----- | ----- | 0---- | 0---0 | ----- | 0---- | 0---0 | ----0 | ---1- | | ----- | ----- | ----- |
| **NLPD** | ----- | ----- | ----- | ----- | 0--00 | 0---- | --1-- | ----- | 0---- | 0---0 | --1-- | 0---- | 0-1-0 | ---00 | ----- | ----- | | --1-- | ----- |
| **VMAF** | --0-- | --0-- | --0-0 | --0-- | --0-0 | --0-0 | ----- | ----- | ----- | --0-0 | ----- | --0-- | 0---0 | ----0 | ---1- | ----- | --0-- | | --0-- |
| **FVVDP** | 1---- | ----- | 1---- | ----- | ---00 | ----- | --1-- | ----- | ----- | ----0 | --1-- | ----- | 0-1-0 | ---00 | ----- | ----- | ----- | --1-- | |

**TABLE IX. Statistical significance groups of metrics for synthetic scenes, real scenes, real scenes with disparity compensation, all scenes, and all scenes with disparity compensation.**

| Synthetic scenes | Real scenes | Real scenes* | All scenes | All scenes* |
|---|---|---|---|---|
| PSNR-YUV | ST-LPIPS | VMAF | DISTS | IW-SSIM |
| MSE-RGB | DISTS | VSI | FVVDP | VMAF |
| PSNR-Y | LPIPS | IW-SSIM | PSNR-HVS | VSI |
| PSNR-HVS | FVVDP | LPIPS | NLPD | VIF |
| NLPD | VMAF | DISTS | MAD | DISTS |
| DISTS | VSI | VIF | PSNR-Y | MSE-RGB |
| GMSD | MSE-RGB | VIFp | IW-SSIM | PSNR-HVS |
| VSI | MAD | ST-LPIPS | GMSD | MAD |
| VIF | GMSD | GMSD | VIF | GMSD |
| VMAF | PSNR-HVS | MAD | VSI | NLPD |
| FVVDP | NLPD | FSIM | PSNR-YUV | FVVDP |
| IW-SSIM | MS-SSIM | MS-SSIM | VMAF | PSNR-Y |
| ST-LPIPS | IW-SSIM | MSE-RGB | ST-LPIPS | VIFp |
| VIFp | VIF | NLPD | FSIM | PSNR-YUV |
| MS-SSIM | PSNR-Y | PSNR-HVS | MSE-RGB | MS-SSIM |
| MAD | VIFp | FVVDP | MS-SSIM | ST-LPIPS |
| FSIM | PSNR-YUV | SSIM | VIFp | FSIM |
| SSIM | FSIM | PSNR-Y | LPIPS | LPIPS |
| LPIPS | SSIM | PSNR-YUV | SSIM | SSIM |

*After disparity compensation

## APPENDIX B
## SA-ST CURVES OF PWRC ON THE NVS-QA DATABASE

Based on PWRC, which has been shown to be reliable in assessing objective quality metrics [55], the SA-ST curves were plotted to compare the performance between the metrics further. The SA-ST plots are shown in Fig. 8. For a given ST value, a higher SA value indicates that the metric is more consistent with human visual perception. As ST increases, the SA of all quality metrics converges to zero, meaning that there are no comprehensible differences in the metrics performances. As presented in Fig. 8, for the overall scenes, FovVideoVDP manages to outperform the DISTS performance, and, with disparity compensation, IW-SSIM is the best metric. For real scenes, both with and without disparity compensation, VMAF consistently reaches the highest values, followed by DISTS and VSI, respectively. Regarding synthetic scenes, metrics based on pixel-wise differences (namely, MSE-RGB, PSNR-Y, PSNR-YUV, and PSNR-HVS) obtain the best SA-ST curves.

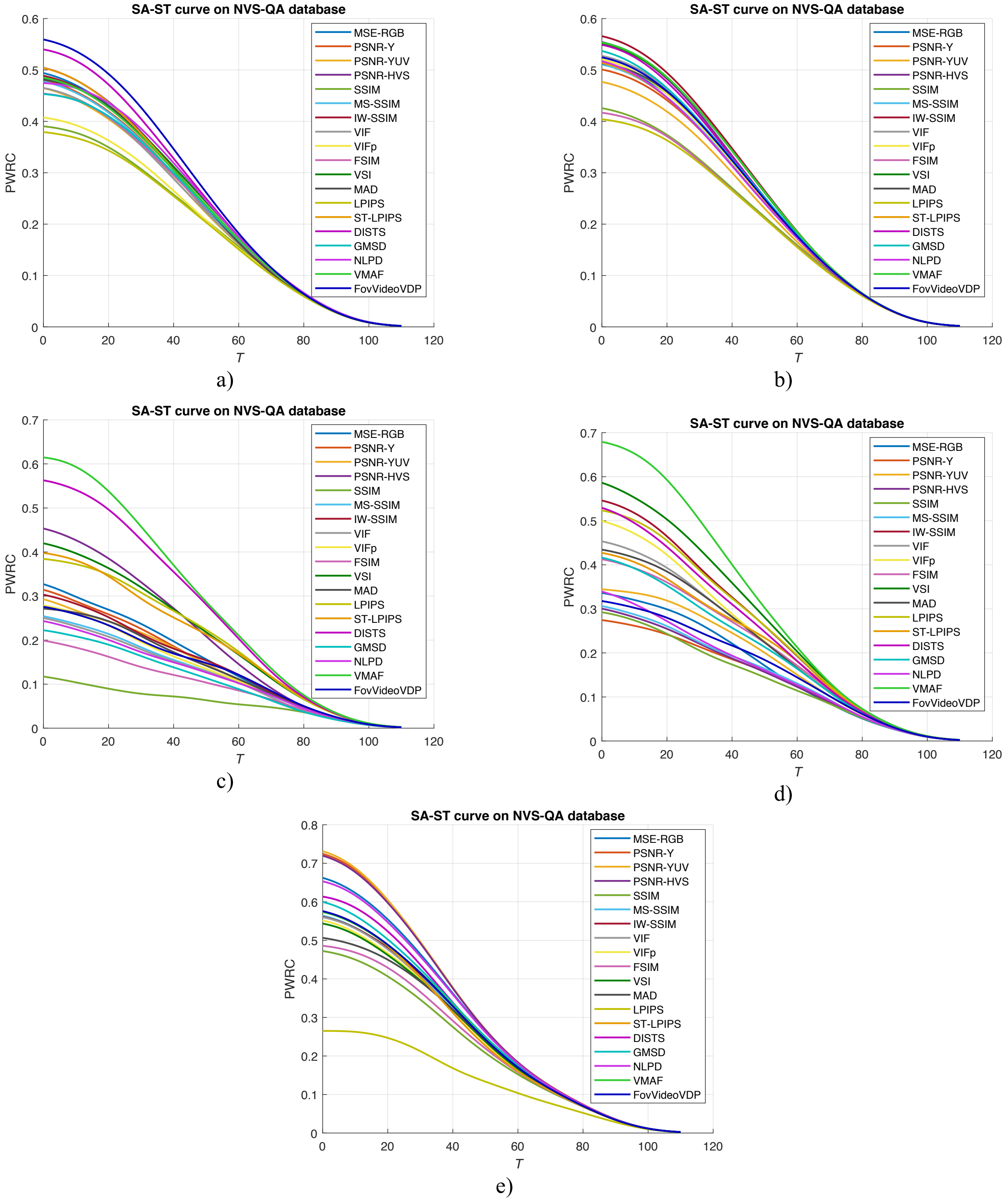

**FIGURE 8. SA-ST curves of the PWRC indicator on ranking the NVS-QA database for a) overall scenes, b) overall compensated scenes, c) real scenes, d) real compensated scenes, and e) synthetic scenes.**

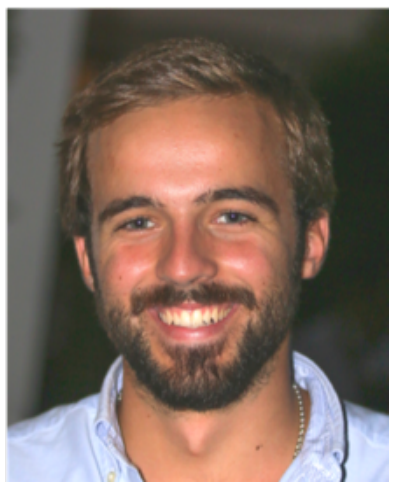

**PEDRO MARTIN** received the B.S. and M.S. degrees in electrical and computer engineering from Instituto Superior Técnico (IST), University of Lisbon (UL), Portugal, in 2019 and 2021, respectively. He is currently pursuing the Ph.D. degree with the Department of Electrical and Computer Engineering, IST/UL. He has been a Researcher with Instituto de Telecomunicações, since 2021, and a Teaching Assistant with the Department of Electrical and Computer Engineering, IST/UL, since 2020. His main research interests include visual quality assessment and coding, with particular focus on neural radiance fields.

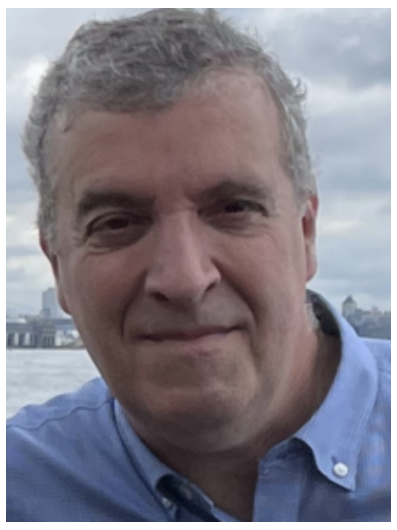

**ANTÓNIO RODRIGUES** (Member, IEEE) received the B.S. and M.S. degrees in electrical and computer engineering from Instituto Superior Técnico (IST), Technical University of Lisbon, Lisbon, Portugal, in 1985 and 1989, respectively, and the Ph.D. degree from the Catholic University of Louvain, Louvain-la-Neuve, Belgium, in 1997. Since 1985, he has been with the Department of Electrical and Computer Engineering, IST, where he is currently an Associate Professor. He is also a Senior Research Member of Instituto de Telecomunicações, Lisbon. His current research interests include mobile and satellite communications, wireless networks, modulation, coding, and multiple access techniques.

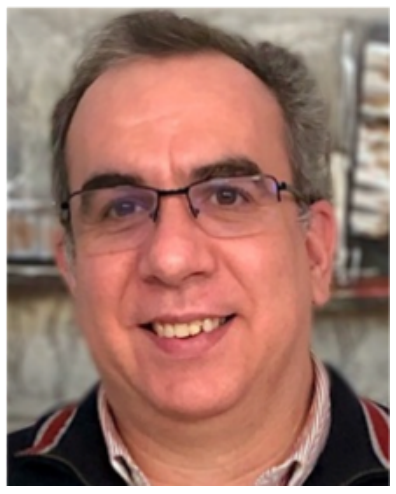

**JOÃO ASCENSO** (Senior Member, IEEE) received the E.E., M.Sc., and Ph.D. degrees in electrical and computer engineering from Instituto Superior Técnico (IST), Universidade Técnica de Lisboa, Lisbon, Portugal, in 1999, 2003, and 2010, respectively. He is currently an Associate Professor with the Department of Electrical and Computer Engineering, IST, and a member of Instituto de Telecomunicações.

He has authored more than 100 papers in international conferences. His current research interests include visual coding, quality assessment, light field and point cloud processing, and indexing and searching of multimedia content. He was an Associate Editor of IEEE TRANSACTIONS ON IMAGE PROCESSING and IEEE TRANSACTIONS ON MULTIMEDIA.

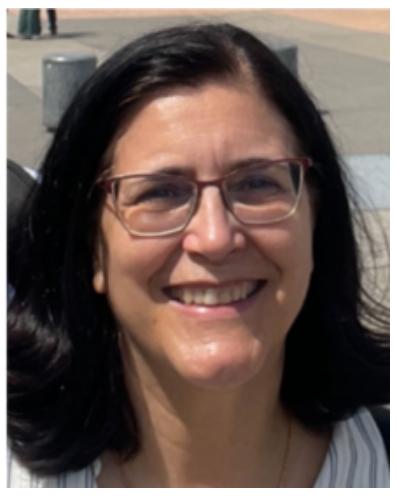

**MARIA PAULA QUELUZ** received the B.S. and M.S. degrees in electrical and computer engineering from Instituto Superior Técnico (IST), University of Lisbon, Portugal, and the Ph.D. degree from the Catholic University of Louvain, Louvain-la-Neuve, Belgium. She is currently an Associate Professor with the Department of Electrical and Computer Engineering, IST, and a Senior Research Member of Instituto de Telecomunicações, Lisbon, Portugal. Her main scientific and research interests include image/video quality assessment, image/video processing, and wireless communications.